\documentclass[12pt]{article}
\usepackage{url} 
\usepackage{hyperref} 
\usepackage{amsmath,amssymb,amsfonts}
\usepackage{graphicx}
\usepackage{hyperref}
\usepackage{geometry}
\begin{document}

\title{The Continuous Logarithm in the Complex Circle for Post-Quantum Cryptographic Algorithms}
\author{Jaafar Gaber \\
\textit{University of Marie and Louis Pasteur, UTBM, CNRS, FEMTO-ST Institute,} \\
\textit{F-90010 Belfort, France}  \\
\textit{Email: gaber@utbm.fr}}
\date{}
\maketitle

\begin{abstract}
This paper introduces a novel cryptographic approach based on the continuous logarithm in the complex circle, designed to address the challenges posed by quantum computing. By leveraging its multi-valued and spectral properties, this framework enables the reintroduction of classical algorithms (DH, ECDSA, ElGamal, EC) and elliptic curve variants into the post-quantum landscape. Transitioning from classical or elliptic algebraic structures to the geometric and spectral properties of the complex circle, we propose a robust and adaptable foundation for post-quantum cryptography.
\end{abstract}

\section{Introduction}
The foundations of classical cryptography are built on hard-to-invert mathematical problems, such as the discrete logarithm and integer factorization. These problems form the basis of many widely used algorithms, including Diffie-Hellman (DH) \cite{DiffieHellman1976}, ECDSA, ElGamal, and elliptic curves (EC) \cite{Koblitz1987}. However, the advent of quantum computers poses a significant threat to these cryptographic systems. Algorithms like Shor's \cite{Shor1994} enable quantum systems to efficiently solve both the discrete logarithm and integer factorization problems, thereby undermining the security of these protocols.

In response to these challenges, we propose a novel cryptographic approach based on the roots of unity and the continuous logarithm in the complex circle. By leveraging the geometric and spectral properties of this framework, our method offers a robust foundation for adapting classical cryptographic algorithms to a post-quantum era. This approach not only preserves key principles of traditional systems but also introduces new structures resistant to quantum attacks, paving the way for future advancements in cryptographic design.

\section{Cryptography Based on the Continuous Logarithm and the Complex Circle}

The continuous logarithm, defined in the complex circle, offers a compelling mathematical basis for designing cryptographic algorithms resistant to quantum attacks.

\subsection{Use of Roots of Unity}
The $n$-th roots of unity, defined as the solutions to $z^n = 1$, are given by:

\[ r_k = e^{i\frac{2\pi k}{n}}, \quad k \in \{0, 1, \dots, n-1\}. \]

These roots form a cyclic group in the complex circle, where $r_1 = e^{i\frac{2\pi}{n}}$ is a primitive root of unity that generates the entire cyclic group. Every element of the group can be expressed as a power of $r_1$, which is essential for ensuring the algebraic structure necessary for cryptographic security. The continuous logarithm is defined by:
\begin{equation}
    L(r^k) = i\frac{2\pi k}{n}.
\end{equation}
This structure provides a robust foundation for cryptographic protocols, particularly for key exchange and digital signature schemes, ensuring resilience against quantum attacks.

\subsection{Cryptographic Properties}

Consider \( Q = r^k = e^{i\frac{2\pi k}{n}} \), where \( r \) is a primitive generator of the group of \( n \)-th roots of unity. The difficulty of the problem lies in solving the continuous logarithm, which involves inverting the exponential operation in this geometric framework. Recovering \( k \) amounts to solving \( L(Q) = i\frac{2\pi k}{n} \), where \( Q \in S^1 \), the unit circle representing the multiplicative group of complex numbers with modulus 1.

More specifically, the difficulty arises from the following properties:
\begin{itemize}
    \item \textbf{Multi-Valued Nature}: The continuous logarithm is inherently multi-valued and defined by $L(Q) = i(\theta + 2k\pi), \ k \in \mathbb{Z}$, where $\theta$ is the principal argument of $Q$. This property introduces an infinite number of possible solutions, increasing the complexity for an attacker attempting to invert the calculation.

    \item \textbf{Angular Ambiguity}: Due to its multi-valued nature, the exact position of $Q$ on the complex circle is ambiguous. This ambiguity complicates the unique identification of $k$, making inversion attacks more challenging.

    \item \textbf{Absence of Exploitable Periodicity}: Unlike the discrete logarithm, the complex circle does not exhibit discrete periodicity exploitable by quantum algorithms such as Shor's, enhancing security.

    \item \textbf{Accumulation of Numerical Errors}: Numerical and angular errors can accumulate during the reconstruction of $k$, particularly for large values of $n$ or in scenarios involving high computational precision or noisy environments.

    \item \textbf{Geometric Dependence}: The reconstruction of $k$ relies on resolving angular ambiguities and managing accumulated errors, embedding the problem within the unique geometric framework of the complex circle.
\end{itemize}

This complexity makes the continuous logarithm problem in the complex circle difficult to solve, even with quantum algorithms. Although this method exploits $n$-th roots of unity in a cyclic framework, more complex generators or hybrid extensions could be considered to further enhance security and expand cryptographic applications (e.g., generators depending on dynamic parameters or combinations with matrix structures).

\section{Post-Quantum Reformulation and Adaptation of Classical Algorithms}

In classical cryptographic algorithms \cite{Menezes1996, Mel2000}, whether in a multiplicative group modulo a large $p$ ($\mathbb{Z}_p^*$, where $r$ is a group generator) or in a group of points on a secure elliptic curve ($E(\mathbb{F}_q)$, where $r$ is a base point on the curve), the difficulty lies in recovering the exponent $k$ in expressions such as $r^k \mod p$ or $k \cdot r$, which corresponds to solving the discrete logarithm problem. This problem underpins many classical cryptographic systems, including:
\begin{itemize}
    \item \textbf{ECDSA (Elliptic Curve Digital Signature Algorithm)}: Uses elliptic curve properties to ensure the security of digital signatures.
    \item \textbf{Diffie-Hellman (DH)}: Enables secure key exchange based on the discrete logarithm problem.
    \item \textbf{ElGamal}: Implements an encryption scheme also based on the discrete logarithm.
\end{itemize}

The geometric properties of elliptic curves \cite{Koblitz1987}, widely used in cryptography, find a natural analogy in the periodic and cyclical properties of the complex circle. This transition preserves geometric foundations while offering a framework better suited to the challenges of post-quantum cryptography.

In the context of the complex circle of modulus 1, where $r$ is a root of unity chosen as a generator, the mathematical foundations change while retaining similar principles. The roots $r^k = e^{i\frac{2\pi k}{n}}$ form a cyclic group under complex multiplication, and the difficulty of recovering $k$ from $Q = r^k P$ (where \( P \) is a fixed base point) relies on solving the continuous logarithm.

This reformulation paves the way for adapting classical algorithms to a robust environment resistant to quantum attacks:
\begin{itemize}
    \item \textbf{Diffie-Hellman (DH)}: The shared key $S = r^{ab}$ relies on the continuous logarithm, strengthening resistance to quantum attacks.
    \item \textbf{ECDSA}: Digital signatures leverage the inherent ambiguity of the continuous logarithm to enhance robustness.
    \item \textbf{ElGamal}: ElGamal relies on the discrete logarithm's difficulty and can be reformulated in this framework. Key generation, encryption, and decryption operations adapt naturally while benefiting from the geometric properties of the complex circle.
\end{itemize}

\section{Conclusion and Perspectives}
This work introduces a new cryptographic approach based on the continuous logarithm in the complex circle. This allows for increased resistance to quantum attacks and enables the reformulation of classical algorithms such as Diffie-Hellman, ECDSA, and ElGamal in a post-quantum framework.

A future direction is to complexify the generator $r$ to further enhance security. For instance, a generator dependent on additional functions or based on roots of complex polynomials could introduce additional algebraic richness. These extensions aim to make the structure of the cyclic group more unpredictable and resistant to attacks while exploring new cryptographic applications.

On the theoretical level, the roots $r^k$ can be modeled by unitary operators in a Hilbert space $\mathcal{H}$, where their behavior can be analyzed using spectral decomposition. This rigorous formalization of the continuous logarithm as a linear operator offers a promising framework for exploring the security of algorithms in advanced post-quantum contexts.

Future work will also focus on studying new cryptographic schemes, including homomorphic protocols based on the complex circle. Homomorphic protocols would ensure secure data processing in sensitive environments while leveraging the geometric structure of the complex circle. Particular attention will also be given to optimizing spectral computations to improve efficiency in high dimensions and to an in-depth experimental analysis of resistance to classical and quantum attacks.


\end{document}